# On the Structural Optimization of Industrial Control Systems Using Metaheuristic Algorithms

Ruslan Zakirzyanov
*NEXT engineering, LLC*
Kazan, Russia
zr@nexteng.ru

***Abstract*—The paper addresses the synthesis of a cost-optimal hardware architecture of the automated process control system (APCS) for continuous processes. The problem is formalized as the search for a minimum-cost rooted tree of devices subject to constraints on the number of I/O channels, the types and number of communication interfaces, the information capacity of processors, the regulation time and the reliability. The signals are grouped by control rooms. The problem is NP-hard, and a hybrid method is proposed that combines a deterministic hierarchy construction with a modified ant colony optimization (ACO) for device-type selection, enhanced by an adaptive heuristic. On ten instances the method is compared with a genetic algorithm (GA), particle swarm optimization (PSO), simulated annealing (SA) and CP-SAT over 30 runs each. By the best found cost the hybrid ACO outperforms GA, PSO and SA on all instances with more than 100 IOs (by 14-38 %) and runs 9-14 times faster. By the median cost, however, the deterministic-greedy baselines are more reliable – GA attains the best Friedman rank – because the cost of the ACO runs is highly variable. All differences are confirmed by the Friedman and Wilcoxon tests. The method targets the early-design search for the cheapest feasible architecture and synthesizes one within minutes at industrial scale (A up to 1086).**

## I. Introduction

Modern distributed automated process control systems of large continuous production facilities in the chemical, petrochemical and oil-and-gas industries are built on control systems integrating hundreds and thousands of technical devices: programmable logic controllers (PLCs), input/output modules, network switches and gateways [1, 2]. The cost, the response time and the reliability of the system are largely predetermined by its hardware architecture, chosen at the early design stages.

In practice the hardware architecture of the control system is formed empirically, relying on the designer's experience and on manufacturer recommendations, which does not guarantee optimality and leads to excessive costs [3]. It also makes an objective comparison of alternative architectures difficult, as cost, reliability and response time are interdependent.

The synthesis of optimal hierarchical architectures has been studied in the theory of organizational and technical systems [4-6], where fundamental results for tree hierarchies were obtained. Applied methods of structural optimization of distributed control systems and production complexes were considered in [7-10], including approaches based on simulation modeling [9] and machine learning [10]. Since the problem is NP-hard combinatorial optimization with multiple constraints, metaheuristics [11-13] are a natural choice. They do not guarantee a global optimum but provide high-quality solutions in acceptable time. For problems naturally represented as a graph, ant colony optimization (ACO) [14] is especially convenient and combines well with local search [15]. This algorithm has been applied to the knapsack problem [16], topology optimization [17], chemical-process optimization [18], parametric [19] and routing [20] problems.

Metaheuristics form a broad family of evolutionary and swarm algorithms, local-search methods and their hybrids [11-13]. Their advantage is finding high-quality solutions in polynomial time and their drawback is sensitivity to the solution representation and to parameter tuning. For problems whose structure is naturally represented as a graph, genetic algorithms are less convenient, as they encode solutions of fixed dimension, whereas ant algorithms construct a solution incrementally, matching the synthesis of a tree hierarchy.

In [21] criteria for selecting the structure of a distributed APCS were proposed and the applicability of metaheuristics was outlined. The present work goes beyond by contributing a refined mathematical model that jointly accounts for channel capacity, interface types, processor information capacity and reliability, a hybrid deterministic-plus-ACO synthesis method with an adaptive heuristic, and a systematic comparison with GA, PSO, SA and the exact CP-SAT method. Most prior approaches either rely on simplified models that do not jointly capture these constraints or are limited to small dimensions.

## II. Problem Statement

The architecture is represented by a rooted tree $G = (V, \mathrm{E})$ whose nodes $V$ are device instances and whose directed edges $E$ define parent-child connections (Fig. 1). The tree height equals $S$: the root is on level 1, leaves on level $S$. A leaf connects field signals, while a single processor on each leaf-to-root path runs the control program.

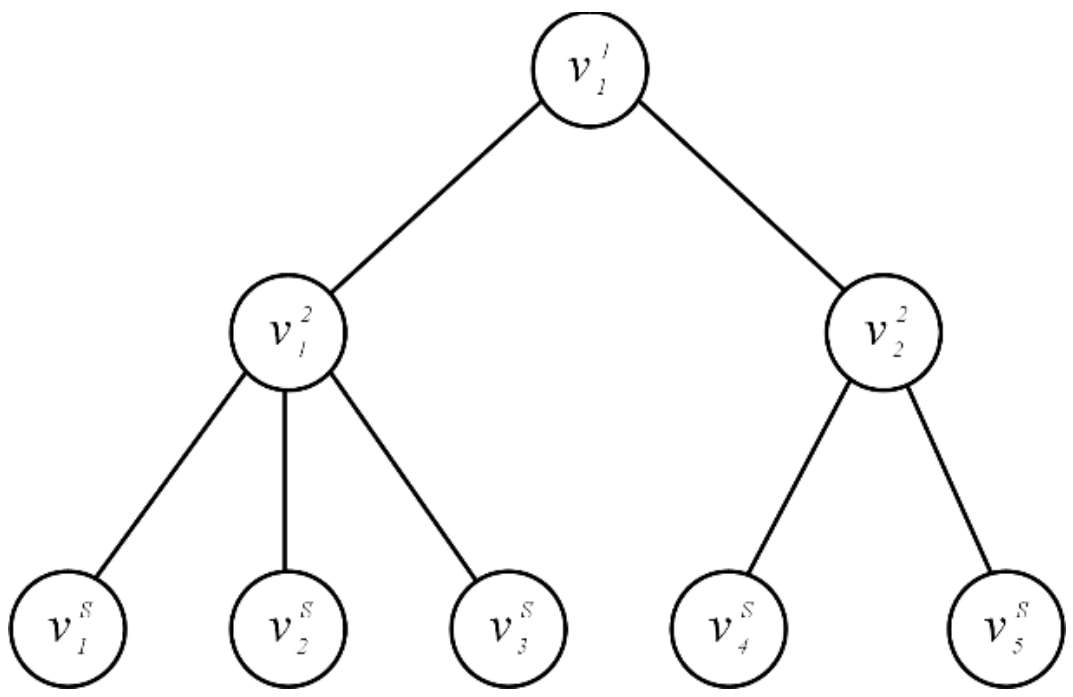

Fig. 1. Hierarchical architecture of the control system

The set of signal and interface types are

$$B = \{1, \ldots, B\}, \quad \mathrm{F} = \{1, \ldots, F\}, \tag{1}$$

where signal type $b$ has a direction $dir(B)$ in $\{0,1\}$ (1 – input, 0 – output), and interface type $k$ has a polling cycle time $tc(k)$. The control rooms form $H = \{1, \ldots, H\}$. Each signal is a pair

$$a = (type(a), croom\ (a)), \tag{2}$$

where $type(a) \in B$ and $croom(a) \in H$. Each device type is described by

$$u_i = (C, P, K, y, T, \gamma, M, N), \tag{3}$$

where $C$ is cost; $P$ failure probability; $K$ the processing capacity; $y$ in $\{0,1\}$ the role (1 – processor, 0 – relay); $T$ the transmission delay (for a processor, the cycle duration); $\gamma$ the parent interface type; $M(f)$ the limits on child devices per interface type *f*; $N(b)$ the number of type-*b* signals connectable at a leaf. Here $P$ is a device failure probability.

With Boolean variables $x(v,a)$, $z(v,a)$, $g(v,h)$ (signal-to-leaf connection, signal processing, leaf-to-room assignment), the model is as follows. Each signal connects to one leaf, and the type-*b* count on a leaf is bounded by its capacity:

$$\sum_{v \in L} x_{va} = 1, \tag{4}$$

$$\sum_{a: type(a)=b} x_{va} \leq N_{u(v),b}, \tag{5}$$

A leaf belongs to one room and $x(v,a) \leq g\big(v, croom(a)\big)$; the child count per interface type $f$ is bounded by $|D(f,v)| \leq M(u(v), f)$. Each signal is processed by one processor within its subtree, and each leaf-to-root path has exactly one processor:

$$\sum_{v \in V} z_{va} = 1,\ \sum_{v' \in P(v)} y_{u(v')} = 1, \tag{6}$$

The processor signal count and the regulation time are bounded:

$$\sum_{a \in A} z_{va} \leq K_{u(v')}, \tag{7}$$

$$T_{reg}(v) = T_{in} + T_{p,in} + T + T_{p,out} + T_{out} \leq T_{max}, \tag{8}$$

where $T_{in}$, $T_{out}$ are the maximum (over input and output signals) sums of the delays $T$ of the devices on the path from the connection leaf to the processor; the leaves attaining these maxima are the anchor leaves. The polling times $T_{p,in}$, $T_{p,out}$ are computed for the anchor leaf as the product of the interface polling cycle time $T_c$ and the number of leaves in the subtree of the corresponding processor port. This form of $T_{reg}$ reflects the real dynamics of the closed loop sensor-processor-actuator in a network with sequential device polling.

Assuming independent failures, the system failure probability [22, 23] is

$$P_{sys} = 1 - \prod_{v \in V}\big(1 - P_{u(v)}\big) \leq P_{max}. \tag{9}$$

The objective is the total device cost:

$$\min_G \sum_{v \in V} C_{u(v)}. \tag{10}$$

Each processor serves signals of at most one room (signals of one room may be split among processors, but not vice versa). Problem (10) subject to (4)-(9) is NP-hard, reducing under mild assumptions to exact cover. The model is built for the preliminary-design stage: a height-homogeneous hierarchy without horizontal links between nodes of the same level, an unlimited number of devices of each type, independent node failures, and device parameters (cost, channel capacity, number and types of interfaces, delays, reliability) taken from a catalog of commercial equipment. These assumptions do not narrow the practical applicability of the model and can be relaxed in further research.

## III. Solution Method

The hybrid method separates synthesis into a deterministic and a stochastic part. The deterministic procedure builds the tree trunk bottom-up and greedily connects the signals grouped by rooms and types, adding new leaves, intermediate nodes and processors as needed while keeping exactly one processor per path. If the set of admissible types is empty or the reliability constraint is violated, the attempt is infeasible and the construction restarts. The stochastic part – a modified ACO – selects only the type of each added device.

Each ant step selects one device type; a path is a chain of such selections building a feasible tree. The type $u_i$ is chosen with probability

$$p(u_i) = \frac{\tau(key)_i^{\alpha} \cdot \eta_i^{\beta}}{\sum_{k=1}^{U} \tau_k^{\alpha} \eta_k^{\beta}}, \tag{11}$$

where $\tau(key)_i$ is the pheromone of the action select type $u_i$ in context key, $\eta_i$ the heuristic desirability, and $\alpha$, $\beta$ are parameters. The context key encodes the stage, level, required interface type, room, signal type, presence of a processor on the path and the role of the lower node. The pheromone is updated only for feasible ants, whereas unsuccessful attempts – a dead end or a violation of the reliability constraint – leave it unchanged. The adaptive heuristic $\eta_i$ dynamically accounts for the level, the number of available channels and the branching possibilities, which raises the share of feasible solutions compared with a myopic cost-based heuristic; the reference solution found by ACO can be refined by local search.

In more detail, the deterministic procedure first builds the trunk: a device admitting signal connection is placed on the bottom level $S$, after which on each level a device with a suitable free port is selected up to the root. The signals are then connected greedily by rooms and types; a new leaf is added when the channels are exhausted, and a new processor (with intermediate nodes built up to the connection point) when the capacity or regulation time is exceeded. The procedure is summarized in Algorithm 1.

**Algorithm 1. Hybrid deterministic + ACO synthesis.**

```
Input: signals by room; device catalog; P_max, T_max
for it = 1..N_iter do
  for ant = 1..N_ants do
    build trunk bottom-up (level S -> root)
    for each device to add:
      pick its type by (11) [pheromone, heuristic]
    attach signals greedily by room and type;
    add leaf if channels full; add processor if
      capacity K or time T_reg (8) exceeded
    if no admissible type or (9) violated:
      mark ant infeasible
  evaporate pheromone (rho); reinforce feasible
    ants by 1/cost; keep best feasible tree
(optional) refine best tree by local search
Output: minimum-cost feasible tree.
```

The algorithm is compared with genetic algorithm (GA), particle swarm optimization (PSO) and simulated annealing (SA). In GA an individual encodes a sequence of device-type selections with tournament selection, crossover and mutation; in PSO a particle position is a vector of device-type preferences; in SA a neighbor is obtained by replacing a device type with a decreasing acceptance temperature. As an exact method, the CP-SAT model (OR-Tools) is used with constraints (4)-(9) in linear form and a 600 s time limit.

For reproducibility, every stochastic algorithm is run 30 times per instance with seeds 1...30. The hybrid ACO uses $\alpha = 2$, $\beta = 1$, evaporation $\rho = 0.25$, $q = 1$ and $\tau_0 = 1$, with the number of iterations and ants set per instance. The baselines use their default settings: GA – a population of 30 over 30 generations, 3 elites, tournament size 3 and mutation probability 0.3; PSO – a swarm of 30 over 30 iterations with inertia $w = 0.5$ and $c_1 = c_2 = 1.5$; SA – 200 cooling steps with factor 0.95, five neighbors per temperature and an auto-calibrated initial temperature; each makes up to 40 attempts to build a feasible initial tree. All algorithms share the same constraint checker, cost function and equipment catalog, so that only the type-selection strategy differs; the full configuration is reproducible from the software [24]. The hyperparameters $\alpha$, $\beta$ and $\rho$ can be tuned further by Bayesian optimization and the multi-objective NSGA-II, which is treated as a separate problem.

## IV. Scientific Novelty

The novelty is threefold. First, a model of a distributed APCS that jointly accounts for constraints on the structure, information capacity, reliability, number of I/O channels and types of communication interfaces, with signals grouped by control rooms. Second, a hybrid method combining deterministic hierarchy construction with metaheuristic device-type selection by ACO. Third, an adaptive heuristic dynamically accounting for the hierarchy level, the number of available channels and the branching possibilities. Together they yield, for industrial-scale objects, the cheapest feasible architecture by the best found cost.

## V. Computational Experiments

The experiments use ten instances: seven synthetic ($A$ = *9...800*) and three real ones from a sulfuric-acid production APCS project ($A$ = *1063*, *1067*, *1086*). Synthetic instances are generated by sampling, for each control room, the number, types and directions of the field signals; all algorithms draw device types from the same catalog. Each stochastic algorithm is run 30 times; the best, mean and median cost, the runtime and the feasibility are recorded.

The comparison with the exact method (Table I) shows that CP-SAT finds solutions only for $A \leq 200$. On $A$ = *9* it proves the optimum (cost 790) and slightly beats ACO (805); as the size grows it hits the 600 s limit and returns a non-optimized feasible solution, and for $A \geq 400$ the linearized model alone needs over 10 GB (MODEL_TOO_LARGE), as it produces millions of Boolean variables. Thus this particular CP-SAT formulation is unsuitable for industrial-scale synthesis; tighter encodings or decomposition might raise the threshold, but exactness at A of several hundred signals remains out of reach for off-the-shelf solvers.

TABLE I. Comparison of CP-SAT and the hybrid ACO

| A | CP-SAT status | CP-SAT cost | ACO cost | ACO t, s |
|---|---|---|---|---|
| 9 | OPTIMAL | 790 | 805 | 1.6 |
| 46 | FEASIBLE | 1974 | 1642 | 9.7 |
| 100 | FEASIBLE | 1539 | 877 | 10.0 |
| 200 | FEASIBLE | 3150 | 1164 | 38.6 |
| 400 | TOO_LARGE | - | 2958 | 29.0 |
| 600 | TOO_LARGE | - | 3451 | 113.6 |
| 800 | TOO_LARGE | - | 5225 | 138.5 |
| 1063 | TOO_LARGE | - | 4769 | 329.7 |

By the best found cost (Fig. 2, Table II) the hybrid ACO is consistently better than GA, PSO and SA for all $A \geq 100$, with a gain from 14 % ($A$ = *400*) to 38 % ($A$ = *200*): the greedy core of GA/PSO/SA confines the search to a narrow region of suboptimal solutions, whereas ACO explores more broadly and reaches a much lower global minimum. The gain is monotone in neither direction but holds on every instance from $A$ = *100* (877 vs 1188 for GA) *to* $A$ = *800* (5225 vs 6653).

TABLE II. BEST ARCHITECTURE COST BY ALGORITHM (C. U.)

| A | ACO | GA | PSO | SA |
|---|---|---|---|---|
| 100 | 877 | 1188 | 1225 | 1225 |
| 200 | 1164 | 1883 | 1795 | 1703 |
| 400 | 2958 | 3445 | 3554 | 3554 |
| 600 | 3451 | 4682 | 4533 | 4533 |
| 800 | 5225 | 6653 | 6680 | 6680 |

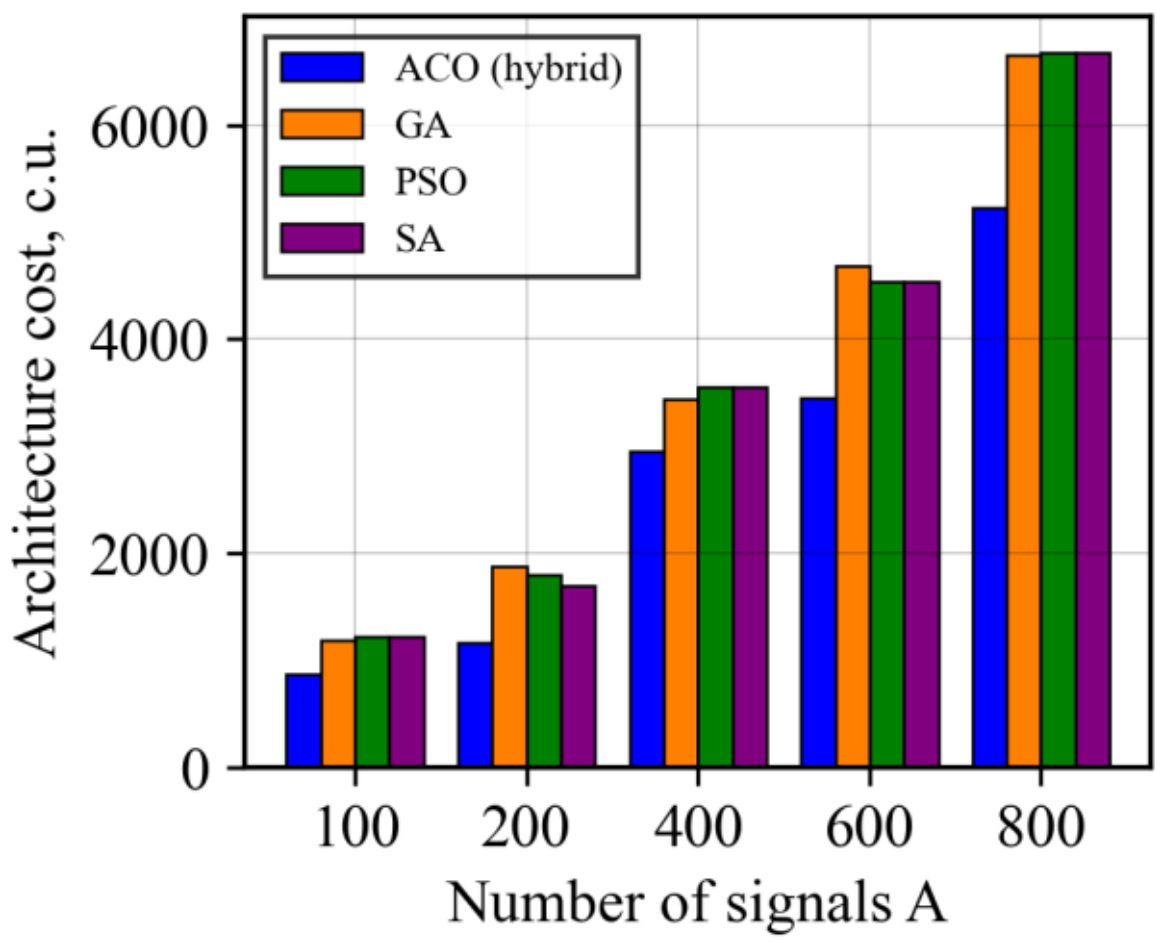

Fig. 2. Best architecture cost by algorithm

In runtime (Fig. 3) the hybrid ACO is markedly faster: the speed-up over GA is *9.9x* at *A = 400* and *13.5x* at *A = 800*, due to the repeated-initialization overhead of GA/PSO/SA, which grows super-linearly with A. A fair engineering criterion is the quality attainable for an equal time budget rather than for an equal number of runs: within the wall-clock time of one GA experiment (30 runs) the hybrid ACO performs from 43 (*A = 200*) to 426 (*A = 800*) runs, so its advantage by the best cost only grows.

A combined cost-times-time indicator (the product of best cost and runtime, smaller is better) summarizes quality and speed jointly: the hybrid ACO outperforms GA by a factor of 3.4 at *A = 100*, 10.8 at *A = 400* and 18.1 at *A = 800*. This indicator, not the runtime alone, is the source of the 18x figure; the pure runtime speed-up stays in the *9-14x* range. The efficiency thus grows with the problem dimension, which is valuable in the main working range of industrial APCS (several hundred to over a thousand signals).

By the median cost the picture differs (Table III). Over five instances (*A = 100...800*) the Friedman test gives $\chi^2 = 13.44$, $p = 0.009$ (Table IV), confirming that the algorithms differ; but the post-hoc mean ranks crown the deterministic-greedy GA (1.40), then SA (2.60), with the hybrid ACO and PSO tied at 3.00 and the trunk-free ACO last (5.00). The significance therefore does not favor ACO by the median: ACO leads the median only on *A = 600* (Fig. 4; paired Wilcoxon vs GA $p < 0.001$, Cliff's $\delta = -0.78$) and trails on the other sizes. The clear, reproducible gain of ACO is in the best found cost (Table II); as a design tool its value is in surfacing the cheapest feasible architecture, which the deterministic baselines never reach. The trunk-free ACO losing on every instance confirms the worth of the hybrid scheme.

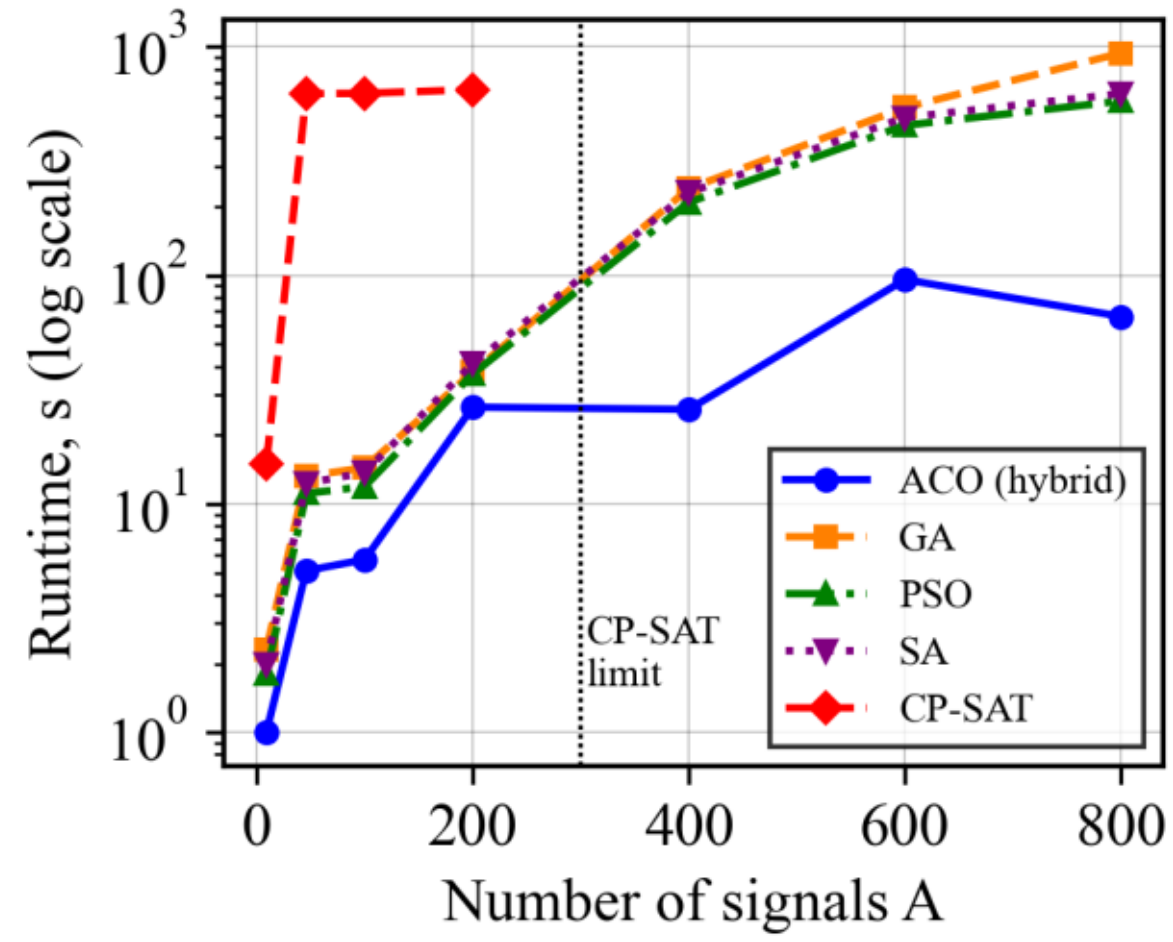

Fig. 3. Algorithm runtime (logarithmic scale)

TABLE III. MEDIAN ARCHITECTURE COST OVER 30 RUNS (C. U.)

| A | ACO | GA | PSO | SA |
|---|---|---|---|---|
| 100 | 1908 | 1325 | 1381 | 1369 |
| 200 | 2600 | 2090 | 2081 | 2117 |
| 400 | 3877 | 3846 | 3938 | 3921 |
| 600 | 4263 | 4975 | 5043 | 5021 |
| 800 | 10482 | 7114 | 7142 | 7115 |

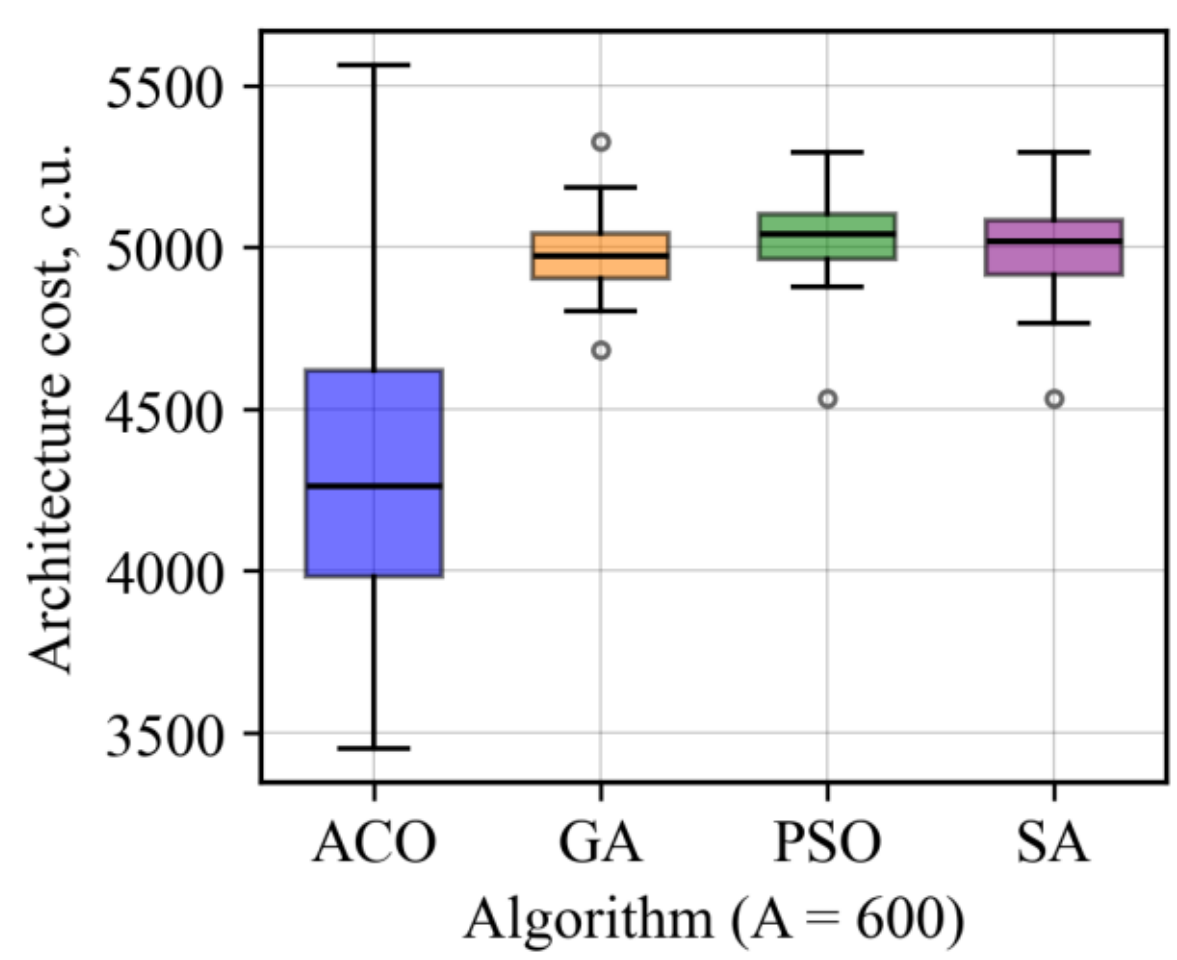

Fig. 4. Distribution of solution costs (A = 600)

The reason the hybrid ACO leads by the best cost yet trails by the median is the spread of its 30 runs: it explores broadly, so the cheapest tree it finds lies far below the deterministic-greedy baselines, but its typical run is dearer and more variable. At *A = 800*, for instance, the ACO cost ranges from 5225 to 11765 (median 10482), whereas GA stays within 6653-7500 (median 7114); GA, PSO and SA, anchored by their greedy core, are tight but cannot reach the cheap regions ACO discovers.

TABLE IV. MEAN RANKS OF THE ALGORITHMS BY MEDIAN COST

| **Algorithm** | **Mean Rank (Friedman)** |
|---|---|
| GA | 1.40 |
| SA | 2.60 |
| ACO (hybrid) | 3.00 |
| PSO | 3.00 |
| ACO (base) | 5.00 |

The practical significance of the method lies in automating the labor-intensive selection of the hardware structure and in enabling a comparison of architecture variants at early design stages under incomplete information about the controlled object. The comparative study covers A up to 800. On the sulfuric-acid project (1063 signals) it produced, within minutes, a feasible architecture (best cost 4769 c. u.) meeting the reliability and regulation-time constraints, based on REGUL R500 controllers.

## VI. CONCLUSION

A hybrid method for synthesizing a cost-optimal hardware architecture of a distributed APCS has been proposed, combining deterministic hierarchy construction with a modified ACO and an adaptive heuristic. Experiments on ten instances showed that the present CP-SAT formulation does not scale ($A \geq 400$), and that by the best found cost the hybrid ACO outperforms GA, PSO and SA on all instances with $A \geq 100$ (by 14-38 %), running 9-14 times faster (up to *18x* on the cost-time indicator). By the median cost, however, the deterministic-greedy baselines remain more reliable, because the cost of the ACO runs is highly variable; this trade-off, confirmed by the Friedman and Wilcoxon tests, is the main limitation to be addressed. The method is implemented as registered software [24].

## REFERENCES

[1] E. L. Itskovich, "Terms of automation and digitalization of process-industry enterprises," Avtomatizatsiya v Promyshlennosti, no. 4, pp. 3-11, 2020 (in Russian).

[2] Yu. S. Tverskoy and A. N. Nikonorov, "On building geographically distributed multifunctional process control systems based on a networked hierarchical control complex," in Proc. XVII Benardos Readings, Ivanovo, 2013, vol. 2, pp. 265-268 (in Russian).

[3] R. M. Zakirzyanov, "Application of metaheuristic algorithms for optimizing the structure of industrial control system," in Proc. VI Int. Conf. Control Tech. Syst. (CTS), Saint Petersburg, Russia, 2025, pp. 34-36.

[4] A. D. Tsvirkun, Fundamentals of the Structure Synthesis of Complex Systems. Moscow, Russia: Nauka, 1982 (in Russian).

[5] A. A. Voronin and S. P. Mishin, Optimal Hierarchical Structures. Moscow, Russia: Trapeznikov Inst. Control Sci. RAS, 2003 (in Russian).

[6] D. A. Novikov and M. V. Gubko, "Methods for optimizing the structure of hierarchical systems," in Large-Scale Systems Control, iss. 2. Moscow, Russia: Fizmatlit, 2015, pp. 359-377 (in Russian).

[7] V. F. Lubentsov, G. V. Masyutina, and E. V. Lubentsova, "Synthesis of the structure of industrial process control systems," Modern Problems and Solutions in Science, Production and Education, no. 6, pp. 9-13, 2018 (in Russian).

[8] A. F. Rezchikov, A. A. Bolshakov, D. Yu. Petrov, and D. D. Yakovlev, "Algorithm for optimizing the structures of robotic assembly complexes," Avtomatizatsiya v Promyshlennosti, no. 5, pp. 3-10, 2025 (in Russian).

[9] I. N. Khrustaleva, D. P. Gasyuk, L. G. Chernykh, and A. A. Laptev, "Selection of a rational structure of the production assembly line based on simulation modeling," in Innovative Ideas in Mechanical Engineering. Saint Petersburg, Russia: POLYTECH-PRESS, 2022, pp. 55-60 (in Russian).

[10] A. H. H. Al Jlibawi, M. L. Othman, A. Ishak, B. S. Moh Noor, and A. H. M. S. Sajitt, "Optimization of distribution control system in oil refinery by applying hybrid machine learning techniques," IEEE Access, vol. 10, pp. 3890-3903, 2022.

[11] C. Blum and A. Roli, "Metaheuristics in combinatorial optimization: Overview and conceptual comparison," ACM Comput. Surv., vol. 35, no. 3, pp. 268-308, 2003.

[12] K. Sorensen, M. Sevaux, and F. Glover, "A history of metaheuristics," in Handbook of Heuristics. Cham, Switzerland: Springer, 2018, pp. 791-808.

[13] L. A. Gladkov, Yu. A. Kravchenko, V. V. Kureychik, and S. I. Rodzin, Intelligent Systems: Models and Methods of Metaheuristic Optimization. Cheboksary, Russia: Sreda, 2024 (in Russian).

[14] M. Dorigo and T. Stutzle, Ant Colony Optimization. Cambridge, MA, USA: MIT Press, 2004.

[15] L. M. Gambardella and M. Dorigo, "An ant colony system hybridized with a new local search for the sequential ordering problem," INFORMS J. Comput., vol. 12, no. 3, pp. 237-255, 2000.

[16] K. Schiff, "Ant colony optimization algorithm for the 0-1 knapsack problem," Czasopismo Techniczne, pp. 39-52, 2014.

[17] Z. Zhang, H. Li, and L. Huang, "The structural topology optimization based on ant colony optimization," Chinese J. Appl. Mech., vol. 28, pp. 226-231, 2011.

[18] V. P. Meshalkin, I. M. Ratkin, and E. V. Petukhova, "Comparative analysis of the efficiency of genetic and ant algorithms for finding the minimum path in optimizing chemical-technological processes," Advances in Chemistry and Chemical Technology, vol. 37, no. 17(279), pp. 58-60, 2023 (in Russian).

[19] Yu. P. Titov and V. A. Sudakov, "Application of the ant colony method to solving parametric problems in the aerospace industry," Trudy MAI, no. 142, 2025 (in Russian).

[20] Y. Wang, P. Chen, Y. Wu, W. Chen, and L. Tan, "Modified continuous ant colony optimization with local search for multiple UAV path planning," in Proc. 6th Int. Conf. Data-driven Optim. Complex Syst. (DOCS), Hangzhou, China, 2024, pp. 557-564.

[21] R. M. Zakirzyanov, "Criteria for selecting the optimal structure of a distributed process control system for large industrial enterprises," Mathematical Methods in Technologies and Engineering, no. 10, pp. 17-23, 2024 (in Russian).

[22] G. P. Akimova and A. V. Solovev, "Methodology for assessing the reliability of hierarchical information systems," Proc. Inst. Syst. Anal. RAS, vol. 23, pp. 18-47, 2006 (in Russian).

[23] M. A. Yastrebenetskiy, Reliability of Automated Process Control Systems. Moscow, Russia: Energoatomizdat, 1989 (in Russian).

[24] R. M. Zakirzyanov, "Program for selecting the optimal structure of a distributed control system," Certificate of state registration of a computer program no. 2025665921, 2025.